# The Regulation of Unlicensed Sub-GHz bands: Are Stronger Restrictions Required for LPWAN-based IoT Success?

David Castells-Rufas, Adrià Galin-Pons, and Jordi Carrabina

*Abstract*—**Radio communications using the unlicensed Sub-GHz bands are expected to play an important role in the deployment of the Internet of Things (IoT). The regulations of the sub-GHz unlicensed bands can affect the deployment of LPWAN networks in a similar way to how they affected the deployment of WLAN networks at the end of the twenty's century. This paper reviews the current regulations and labeling requirements affecting LPWAN-based IoT devices for the most relevant markets worldwide (US, Europe, China, Japan, India, Brazil and Canada) and identify the main roadblocks for massive adaption of the technology.**

**Finally, some suggestions are given to regulators to address the open challenges.**

## I. INTRODUCTION

THERE are different predictions about the number of devices that will become connected to the internet in the near future with the widespread of the Internet of Things concept (IoT). Either being 50 G by 2020 [1], or 75 G by 2025 [2], it seems to be a consensus about the disruptive nature of IoT [3] and about the number of connected devices being in the order of billions.

The confluence of the evolution of many technologies like energy scavenging, machine-to-machine communications, and low power wireless communication technologies support the narrative that any device that would benefit from being connected will definitely be connected since the cost of the connection will be insignificant. This cost includes the cost of the chips, the cost of the communication channel, and the cost of the energy.

However, to the best of our knowledge, the studies in the literature are not so explicit about predicting the number of devices that will be wirelessly connected through low power or low throughput radio communication links. In many works

(like [4][5]) it comes implicit that a high percentage of them will be connected by wireless links, as some of the fundamental technologies enabling IoT are Low Power Wide Area Networks (LPWAN) working on unlicensed bands, which are free to use.

Nevertheless, the use of the radio spectrum is regulated in most countries of the world. This aspect is often overlooked in the literature, not considering the limitations that regulation could impose on the deployment of such technologies. Our hypothesis is that current regulations can hamper the deployment of wireless IoT applications due to their impact on the spectrum use and the microelectronics industries.

The paper is organized as follows: we describe the radio spectrum in Section II, and recall the events that shaped the current spectrum regulations in Section III. Section IV presents the different technologies in use in the IoT wireless landscape. Section V reviews the regulation and certification process on the main world markets. In Section VI we analyze what rules the regulators can enforce in trying to orchestrate the spectrum. In Section VII we analyze the mathematical expressions that could describe the node density and bitrate density of LPWAN. In Section VIII we estimate the maximum values for LoRa and Sigfox technologies given on the scope of different regulations. Those results are contrasted with the results from the literature in Section IX. In Section X we study additional economic impacts caused by the current regulation. In Section XI, before concluding, we discuss the benefits of the harmonization of regulations.

## II. RADIO-SPECTRUM LIMITS

Although, theoretically, the radio spectrum is an infinite resource, the interesting frequency bands for communication over the earth surface are delimited by two factors:

1) In the low end, by the Shannon-Hartley theorem (Eq. 1), which relates the amount of information potentially transmitted over a channel.

$$C = B \log_2\left(1 + \frac{S}{N}\right) \tag{1}$$

Where $C$ is the traffic capacity of the channel in bits per second, $B$ is the bandwidth of the channel in Hertz and $S/N$ is the signal to noise ratio of the channel. So if we want to

This work has been partly funded by the Serene-IoT project (Penta 16004).

D. Castells-Rufas is with the Microelectronics and Electronic Systems Department, Universitat Autònoma de Barcelona Bellaterra, 08193 Spain (e-mail: david.castells@uab.cat).

A. Galin-Pons is with R&D Department, Applus+ Laboratories, Bellaterra, 08193 Spain (e-mail: adriagalinpons@gmail.com).

Jordi Carrabina is with the Microelectronics and Electronic Systems Department, Universitat Autònoma de Barcelona Bellaterra, 08193 Spain (e-mail: jordi.carrabina@uab.cat).

transmit an amount of information in a time period either we use enough bandwidth or we have enough signal to noise ratio. In this trade-off we have a limited ability to increase the signal to noise ratio of radio channels, it is easier to select the carrier frequencies that will provide enough bandwidth to allow the required traffic capacity.

2) In the high end, frequencies above PHz are known to be ionizing radiation and harmful to human life, so they are avoided. Secondly antenna efficiency has an intrinsic attenuation relation with frequency, i.e. a reduction of the received power ($P_{rx}$) with respect to the emitted power ($P_{tx}$). This is known as free space path loss (FSPL). Ignoring the gain effects of both antennas the loss is described by Eq. 2, where $d$ is the distance, $f$ the frequency, and $c$ the speed of light.

$$FSPL = \frac{P_{Tx}}{P_{Rx}} = \left(\frac{4\pi d f}{c}\right)^2 \qquad (2)$$

Moreover, different frequencies propagate differently in the atmosphere. Especially frequencies at the GHz ranges are absorbed by atmospheric gases such as $O_2$, $H_2O$, etc.

Another factor that influences the suitability of different frequencies is the earth curvature, which limits the range of direct line of sight propagation to a distance known as radio horizon. The radio horizon is mainly determined by the height of the communicating antennas.

An alternative propagation medium is the surface of the earth. Ground wave propagation is possible below 3 MHz, but it is practically unfeasible to go further than some hundreds of kilometers.

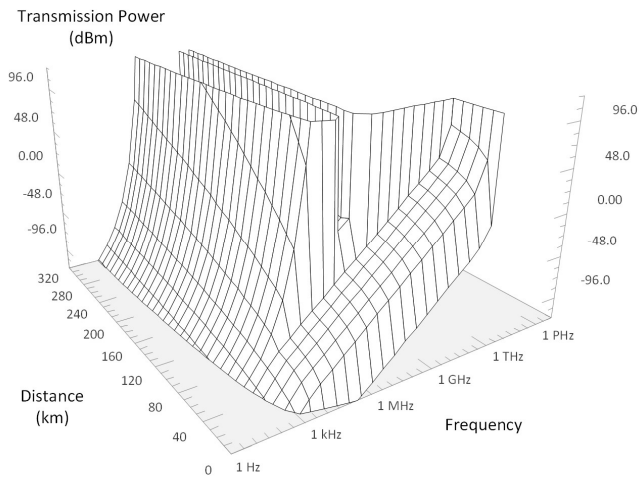

Figure 1 Illustrative simplified example of the transmission power that a transmitter should use so that a receiver endpoint can decode the signal depending on the frequency of the carrier and the distance of the receiving endpoint assuming an antenna height of 200m, and -120 dBm of receiver sensitivity. For lower frequencies, surface wave propagation allows a long distance range. For higher frequencies the propagation is limited to the radio horizon, except in the Skywave band. For extremely high frequencies above 30GHz the absorption of the wave's energy by atmospheric gases limits the transmissions to very short distances.

For certain frequencies and the appropriate atmosphere conditions the ionosphere contributes to allow what is known as Skywave propagation, increasing the possible range to a much longer distance.

Figure 1 depicts a simplified model of the different effects that contribute to the cost of transmitting information from a sender to a receiver using different frequencies and different distances between both endpoints.

This is a simple model with just two endpoints. Current communication systems are typically not so simple, and cost is more complex to compute. The observed limitations have been overcome by deploying networks of antennas and satellite-based communications. In this context, there is not a single efficiency measure, but several, like spectrum efficiency (SE) or energy efficiency (EE) [6].

Nevertheless, since the radio spectrum is a scarce resource, it comes as no surprise that economic laws and policymakers play an important role to orchestrate its exploitation in an attempt to maximize its utility.

## III. Brief History of Radio Spectrum Regulation

Going back in history, after the discovery of the possibility of transmitting information through electromagnetic waves, radio was mostly used for Morse communication, but at the beginning with no regulation. Regulations were later introduced in the Berlin 1903 and London 1912 conventions to orchestrate different international radio services with an important focus on emergency situations.

Shortly after the sinking of the Titanic, US adopted the Radio Act of 1912, taking a leadership position that it would maintain for the rest of the century. The main early beneficiaries of the radio technology were still maritime ships. The International Telecommunication Union (ITU), an International Regulatory Body (IRB) had been founded previously, in 1865. Regional and International Regulatory Bodies (RRB and IRB) were playing an important role to ensure effective communications within different territories. National Regulatory Bodies (NRB) were still not needed because the technology was either controlled by governments or in hands of very few pioneers.

The invention of the amplitude modulation (AM) and its application for voice transmission caused the introduction of commercial broadcast radio stations. Soon after the first commercial radio emission by KDKA in 1920, the number of transmitters, both commercial and amateur, proliferated at a fast pace creating a chaotic situation with thousands of amateur broadcasters and common interferences to commercial radio stations. The US government saw the need of licensing different radio bands and established transmission power limits in the Radio Act US 1927 to solve the situation.

In the following decade many advances were made. Television [7] was improved and Television broadcasters appeared slowly as new users of the radio spectrum. Frequency Modulation [8] was invented as a better alternative to AM thanks to its lower interference features.

In this dynamic scenario the US government issued the Communications Act of 1934, which created the Federal

Communications Commission (FCC), a NRB to regulate the radio spectrum in US. Regulators not only licensed frequencies and regulated transmission power, but also introduced the mandatory use of communication standards in certain licensed frequencies. For instance, in 1941 the FCC created the NTSC standard making it mandatory for the VHF television channels.

As the regulators either limited or licensed parts of the radio spectrum, they raised a conflict with other uses of radio technology that had been discovered in previous decades. In addition to telecommunications, radio could be also used for induction heating, dielectric heating (microwave heating), diathermy, inducing mechanical vibration, ionization of gases, particle acceleration, etc. In order to avoid limiting the advances on those technologies, the Industrial, Scientific and Medical (ISM) bands were first established at the International Telecommunications Conference of the ITU in Atlantic City in 1947, with the aim of allowing some unlicensed bands for those applications to use free of charge. NRBs later adapted the concept introducing some limitations on emitted power and duty cycle.

Initially, it was forbidden to use unlicensed bands for communications. They could exclusively be used for ISM applications. But the advances in electronics and computing caused a big market pressure demanding unlicensed bands to allow short-range wireless communications [9]. At the same time, the market was also demanding permission to benefit from the advances on spread spectrum modulation, which had been invented during the war as a military technique to increase the security of communication channels [10], but remained forbidden for civilian use. The FCC finally allowed communications on the ISM bands and the use of spread spectrum in 1985.

In the new scenario, regulation became very complex and a new problem arose. The risk putting a non-conformant product in the market was high. Again, in order to protect industrial investments, governments decided to mandatorily require the pre-certification of all new products using unlicensed bands. The new regulations were introduced in 1989 under the FCC Part 15 rules [11]. In Europe the European Telecommunication Standard Institute (ETSI) was conceived the previous year in 1988.

Until our days, the following technological advances had not a big impact in the mandatory regulations of sub-GHz unlicensed bands. Nonetheless, the market pressure on the continuous demand of more spectrum drove regulators to start mandating for higher levels of spectrum efficiency. The FCC issued a narrowbanding mandate [12] to migrate VHF/UHF licenses to higher spectrum efficiency systems by the beginning of 2013.

Furthermore, from the certification perspective, private-companies created different associations to promote common technologies. Those associations often provide their own certification program, such as Wi-Fi Alliance Certification Program, Bluetooth SIG, or Sigfox Ready to name a few.

However, there is still some discussions (like shown in [13][14]) whether the spectrum use should be more controlled by free market forces and less driven by the command and control of governments.

## IV. IoT LANDSCAPE

IoT is based on the idea that a myriad of devices will be connected to the Internet. Some examples of these objects could be home appliances, machines, vehicles, or embedded sensors. Their connection will allow the acquisition of new data and the opportunity to create new business models.

It is generally assumed that wired networks will be part of the networking infrastructure but will not provide the access connection to most end devices. One reason is the cost of infrastructure, but another important reason is that wireless networks allow mobility. Without the need of wired communication infrastructure the open challenge for wireless devices is power supply. There are four possible strategies to power such devices: 1) connection to the power grid 2) rechargeable batteries 3) energy scavenging 4) life-long batteries.

The chosen strategy has a big impact on the communication capabilities of such devices. Basically, as seen in Section II, the more power is available, the more bandwidth the device can use.

There are currently several available wireless technologies with different properties and different target applications. Their radio interfaces present multiple trade-offs between relevant parameters which will determine the network behavior, including: latency, mobility, cost, capacity, power consumption, complexity, reliability, interference immunity, symmetrical uplink and downlink channels, etc.

Nevertheless, following the ETSI classification, the IoT landscape can be sorted out in four main groups:

- **Cellular based**: all technologies based on cellular technologies optimized for IoT, including: LTE-CATM, NB-IoT and E-GSM. All this technologies take advantage of the licensed band pros.

- **Dedicated Star Networks**: technologies which its network typology is a star and are optimized for IoT. They are built over shared spectrum: Sigfox, LoRaWAN, Weightless, Telensa, etc.

- **Dedicated Mesh Network**: mesh networks covering wide area with multi-hops connectivity -these systems are also known as Network-Based SRDs in ETSI EN 303 204-. Silverspring technology is an example of Dedicated Mesh Network.

- **Low power versions of LANs & PANs**: Like WiFi, Bluetooth (5.0/4.2/4.1/4.0, Low Energy) , WiGig, Ingenu, ZigBee, Thread, Z-wave, EnOcean, etc. They are also unlicensed technologies however the coverage range is much shorter than the second group presented above.

The first two subgroups above (Cellular and dedicated star networks) were referred to as LPWAN by many analysts. These two types of radio techniques share the common use of high sensitivity for increased radio coverage and the low power consumption.

The term IoT-LTN [15] refers to the Dedicated Star Networks category, which, in addition to the characteristics of LPWAN, adds the properties of shared spectrum, random channelization, star topology and half duplex communication.

Table 1 presents the characteristics of the main IoT related physical layers grouped by ETSI classification. In this work we will ignore the issues with the higher layers in the International Standards Organization (ISO) communication stack. Notice that dedicated star networks work in the sub-GHz bands offering a very low bitrate. They usually use modulation techniques that require less computational power (and energy) than the higher speed networks and can tolerate challenging SNRs so low as -20 dB.

The need for LTN is motivated by the type of devices that are powered by life-long batteries or energy scavenging systems. Those kind devices have a very limited energy budged that cannot be wasted on constant network connection. Moreover, it is well known that with modern modulations receiving is more power hungry than emitting, so this limited power scenario will definitely incentivize (mostly) unidirectional traffic from nodes to gateways that have a wired power supply that allows them to constantly listen to the used radio channels. The need for long range coverage is the result from the economic pressure. Gateways with power supply and Internet connection will usually have a much higher cost than end nodes, so it is desired to amortize their cost on the maximum number of end devices.

Some of the best candidates to take profit of such LTN networks are Wireless Sensor Networks (WSN) [16].

## V. THE APPROVAL PROCESS AROUND THE GLOBE

The management of the radio spectrum has been assumed by NRBs in most states, which implement their desired policies following the agreements made by RRB and IRBs. Nations must report their progress in applying the decisions from the ITU and the World Radiocommunication Conferences (WRC), which try to harmonize global practices.

As a general approach, each target market has its own regulation scheme for introducing a given RF Sub-GHz band technology or generic radio transceiver as well as dedicated certification process which most heavily impacts chip manufacturers/integrators and IoT importers, as they are forced to spend time getting acquainted with the local legal requirements for their devices.

The process, illustrated by Figure 2, starts with the manufacturing of a system, the integration of preexisting parts into a system, or even with the import of a product manufactured abroad. Each product must be tested against a normalized test plan conceived by the regulator. The use of pre-certified modules integrated into the final host product may help to reduce the associated testing costs. In the certification step the results of tests are analyzed together with additional technical documentation. If the process is successful a label is issued, which allows the access to the market.

In some countries local representatives are needed to be able to access the market. This fact could influence the expansion strategy of manufacturers so that they prioritize to invest in facilities where the representatives are mandatory.

TABLE 1
RADIO TECHNOLOGIES FOR THE PHYSICAL LAYER OF WIRELESS INTERNET OF THINGS

| Category | Technology | Governing Body / Standard | Frequency bands | Capacity (kbps) | Multiple Access | Modulation |
|---|---|---|---|---|---|---|
| Cellular based | LTE-CATM | 3GPP Rel 13 | LTE | 1024 | OFDMA | QPSK, 16QAM, 64QAM |
| | NB-IoT | 3GPP Rel 13 | LTE/GSM | 250 | OFDMA | BPSK, QPSK, 16QAM |
| | EG-GSM | 3GPP Rel 13 | GSM | 240 | TDMA | GMSK, 8PSK |
| Dedicated Star Networks | Sigfox | SIGFOX | <1 GHz | 0.6 | UNB/FHSS | GFSK/DBPSK |
| | LoRaWAN | LoRa Alliance | <1 GHz | 50 | CSS | (G)FSK |
| | Weightless-P | Weightless SIG | <1 GHz | 100 | FDMA + TDMA | GMSK, OQPSK |
| | Telensa | WIoTF | <1 GHz | 0.5 | UNB/FHSS | 2FSK |
| Dedicated Mesh Network | Silverspring | Wi-SUN Alliance IEEE 802.15.4 | <1 GHz , 2.4 GHz | 1024 | CSMA/CA | MR-FSK/MR-OFDM/MR-O-QPSK |
| Low power versions of LANs & PANs | WiFi | WiFi Alliance IEEE 802.11a/b/g/n/ac | 2.4 GHz, 5 GHz | 11000-6900000 | OFDM, DSSS, OFDMA | CCK, BPSK, QPSK, 16-QAM, 64-QAM, 256-QAM |
| | Bluetooth (4.0/4.1/4.2 LE) | Bluetooth special interest group (SIG) | 2.4 GHz | 1024 | TDMA | ASK, FSK |
| | Ingenu | Ingenu (formerly OnRamp) | 2.4 GHz | 20 | RPMA | BPSK, OQPSK,FSK, GFSK, P-FSK, P-GFSK |
| | ZigBee | ZigBee Alliance IEEE 802.15.4 | <1 GHz , 2.4 GHz | 250 | CSMA/CA | DSSS, BPSK, O-QPSK |
| | Thread | Thread Group IEEE 802.15.4 | 2.4 GHz | 250 | CSMA/CA | DSSS, O-QPSK |
| | Z-wave | Z-Wave Alliance ITU G.9959 | <1 GHz | 100 | TDMA | FSK, GFSK |
| | EnOcean | EnOcean Alliance ISO/IEC 14543-3-1x | <1 GHz | 125 | TDMA | ASK, FSK |
| | WiGig | WiFi Alliance IEEE 802.11ad | 60 GHz | 6760000 | SC-SS | $\pi$/2-BPSK, QPSK, QAM16, SQPSK, QAM64 |
| | Dash7 | Dash7 Alliance | <1 GHz | 167 | TDMA | (G)FSK |

One of the responsibilities of national authorities is to perform appropriate monitoring and post-market surveillance once the IoT-LTN devices are in the market.

Manufacturers, importers or distributors must bear in mind that, at any moment, national authorities may ask for compliance exhibits. So, it is highly recommended to have always a product sample available. Stating that a device will not be marketed or that is no longer manufactured is not a sufficient justification for not providing post-certification production samples upon request.

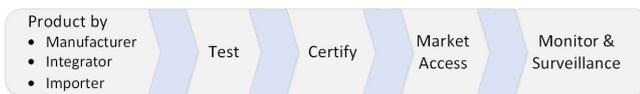

Figure 2 General approval procedure for new radio equipment to gain access to the market.

Although the process is quite similar across the world there are some differences among countries. We outline the relevant details of the process for the main global markets in terms of Gross Domestic Product (GDP) (as obtained from [17]). Those are United States of America, the European Single Market, englobing the European Union (EU) and European Free Trade Association (EFTA) states, People's Republic of China, Japan, Republic of India, Federative Republic of Brazil and Canada.

### A. United States of America

In the United States of America, the communications regulations are set by the FCC together with the National Telecommunications & Information Administration (NTIA); the unlicensed equipment and intentional radiators regulations such as Unlicensed IoT-LTN are present in the 47 CFR FCC Rules Part 15 [18]. Testing versus those requirements shall be performed by a recognized testing laboratory by the FCC.

The certification for equipment subject to the FCC's certification procedures for transmitting devices is handled by a Telecommunication Certification Body (TCB),- a third-party organization which is devoted to review and evaluate the requirements fulfilment and to upload the documentation to the FCC database for approval. There are a number of TCBs distributed around the globe since the FCC rules established procedures for the recognition of foreign TCBs under the terms of a government-to-government Mutual Recognition Agreement/Arrangement (MRA).

### B. Europe

In Europe, the applicable laws are derived from the Directive 2014/53/EU of the European Parliament, which specifies the requirements on Health and Safety, Electromagnetic Compatibility, and Effective use of Radio Spectrum for new products.

There are no specific requirements on who is allowed to do the testing step. Nevertheless, European commission names a list of organizations as Notified Bodies to perform the certification step.

Additionally, the manufacturers can do the certification on their own, and assume the presumption of conformity, if the type of device is covered by any category of the existing standards published on the Official Journal of the European Union. Otherwise, the certification must be done by a Notified Body.

Self-certification can be risky if the manufacturer is not versed in the European standards and the activities held by standardization bodies like ETSI or CENELEC. The applicable requirements for IoT-LTN devices fall under the Short-Range Devices category regulated by ERC Recommendation 70-03 [19].

### C. China

The IoT-LTN applicable standard in China is the SRRC 423 [20] (in traditional Chinese language), which list the required parameters and functions that must be tested for radio transmission equipment. Testing activities shall be carried out by an Accredited Chinese Laboratory. Next, before gaining access to the Chinese market, two certification schemes are required for IoT-LTN products: an approval from the Ministry of Industry & Information Technology (MIIT) and the China Compulsory Certificate (CCC or 3C).

In addition to the typical product certification, which in China's case is issued by MIIT, the Chinese government enforces a certification on the production factories. This is implemented by the CCC certification that involves an audit to the production lines (either in China or abroad) by Chinese accredited authorities.

The market surveillance activities are performed by the State Radio Monitoring and Testing Center (SRTC).

### D. Japan

In Japan, all the approval scheme is set by the Radio Law (Law No. 131 of May 2, 1950) which regulates the general provisions for introducing a given Radio product into the Japanese market, considering the applicable technical requirements, testing and certification schemes.

Certification organizations, known as Registered Certification Bodies (RCB), shall be registered by the Ministry of Internal Affairs and Communications (MIC). MIC regulates the testing procedures for specified radio equipment in Notification No.88 of MIC, 2004. According to the Article 38-2 of the Radio Law, every type of specified radio equipment is tested by RCBs or competent laboratories.

### E. India

The Radio-spectrum regulations in India are driven by the Telecommunications Engineering Center (TEC), a group of the Ministry of Communications of the Indian Government. The Indian Telegraph (Amendment) Rules from 2017, describe the test and certification scheme prior to sale, import or use in India.

The Indian Regulation consists of a collection of essential requirements that a given device shall fulfil. Regarding the IoT-LTN equipment, the corresponding essential requirements are under TEC2449:218. All the testing activity shall be done by Indian Accredited Lab designated by TEC following the Mandatory Testing and Certification of Telecom Equipments (MTCTE).

Once the testing is completed and successfully demonstrated that given device fulfils all essential requirements, the certifications must be carried out by TEC Officers based on test reports and additional technical documentation.

### F. Brazil

The body taking care of the spectrum use and regulations in Brazil is the Agência Nacional de Telecomunicações (ANATEL). All telecommunication products to be used in Brazil must be certified. The Regulation on The Certification and Authorization of Telecommunication Products, approved by Resolution No. 242, of 30 November 2000 establishes the general rules and procedures related to the certification and authorization of telecommunications products.

The testing activity against the local requirements must be carried out by In-country test laboratory properly recognized according to the local requirements stated by ANATEL.

Once the testing is carried out and given device fulfills all applicable technical requirements the certification takes place by ANATEL. A local representative is also required according to the Brazilian certification scheme.

### G. Canada

In Canada, it is the Innovation Science and Economic Development Canada (ISED) that is in charge of the Radio Frequency Regulation and the Radio Standards Specification (RSS).

The "RSS-Gen General Requirements for Compliance of Radio Apparatus Issue 5 (2018)" sets out the general requirements for radio apparatus that are used for radio communication.

Testing laboratories test the products in accordance with the enforced regulations, and certification bodies (CBs). It is possible that third party recognized independent organizations certify the radio-communication equipment.

The Testing Laboratories and Certification Bodies that are recognized by the ISED are listed on the Government of Canada website. The technical requirements for IoT-LTN devices are set on RSS-210.

The responsible party of a given product must be within a Canadian soil address. Foreign entities shall require a local representative in order to start commercial activities in Canada.

A summary of the situation in the different analyzed regions is given in Table 2.

## VI. TECHNICAL CONSTRAINTS DERIVED FROM REGULATIONS

With spectrum management nations usually pursue the maximization of the utility of the spectrum. As we reviewed in section III they started with a "command and control" approach, which has been later adapted to a more market driven approach for certain bands [21]. It is complex to define utility, but in the modern capitalist view of society, it should have some link with a part of a nation's GDP. Following this reasoning, regulators would aim to foster economic activity around the use of the radio spectrum (as shown in [22]). In any case, the job of the regulator is to select the appropriate incentives that encourage the market players to invest their resources to create new wealth.

For unlicensed bands, money is not in the incentives game, so regulators select some technical parameters of radio transmission and decide some arbitrary thresholds following reasoned criteria. The responsibility of analyzing that all products using unlicensed bands are fulfilling the requirements is usually delegated to Certification Bodies (CBs), who verify that all the technical parameters are in the acceptable ranges. Although some monitoring can be done (and should be done) to ensure that all the products are well behaving when deployed in the market, it has a much lower cost for the state to require certification before market access.

TABLE 2
PHY DETAILS OF THE CERTIFICATION PROCESS ACROSS THE TOP TEN GDP COUNTRIES WORLDWIDE

| | US | Europe | China | Japan | India | Brazil | Canada |
|---|---|---|---|---|---|---|---|
| Reference Standard (test) | 47 CFR FCC Rules Part 15 subpart C §15.247 | ETSI EN 300 220-2 EN 303 204 | SRRC 423 | Notification No.88 of MIC ARIB STD - T108 | TEC2449:218 | Resolution No. 242 Resolution No. 506 | RSS-GEN RSS-247 |
| Test Body | Recognized ISO 17025 Lab | Own / Other | Chinese ISO 17025 Lab | Recognized ISO 17025 Lab | Recognized ISO 17025 Lab | Brazilian Recognized ISO 17025 Lab | Recognized ISO 17025 Lab |
| In-country testing required | No | No | Yes | No | Yes | Yes | No |
| Labelling | FCC ID: XXX-YYYYY | 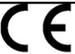 | 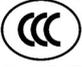 CMIIT ID 2018yznnn | 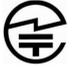 XXX - ABCDEF | 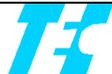 PQRS: ABCDEF | 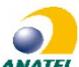 XXXXX-YY-ZZZZZ | ISED ID: XXXXX-YYYYY |
| Certification Body | TCB | Own Producer / Notify Body (DoC if HS or NB UE type examination.) | MIIT | RCB | TEC | OCD | CB |
| Typical Lead Time (Test & Certificaion) | 6 weeks | 4 weeks | 12 weeks | 9 weeks | 9 weeks | 9 weeks | 6 weeks |
| National Local Representative Required | No | No | No | No | Yes | Yes | Yes |

The first decision of the regulator is to select the **frequency bands and its applications**. This is usually done taking into account the effects analyzed in Section II. Higher frequencies are usually used for high throughput networks (see Eq. 1). Since the power required for transmission is positively correlated with the distance and the carrier frequency of endpoints, lower frequencies are often used for longer range communications. For a given band, the regulator can establish a **maximum Tx power limit**, which almost automatically results in determining a maximum coverage radius (see Eq. 2). The probability of interference can be very high if no further rules are enforced, making the use of the band too unpredictable for any successful business model to succeed.

TABLE 3
DETAILS OF THE TECHNICAL CONSTRAINTS BY REGULATIONS FROM THE TOP TEN GDP COUNTRIES WORLDWIDE

| | US | Europe | China | Japan | India | Brazil | Canada |
|---|---|---|---|---|---|---|---|
| **General Parameters** | | | | | | | |
| Frequency Range (MHz) | 902-928 | 863-875.6 | 779-787 | 915.9-916.9 920.5-929.7 | 865-867 | 902-907.5 915-928 | 902-928 |
| Maximum TX Power (dBm) | 30 (>50 ch.[1]) 24 otherwise | 27 (869.4-869.6) 14 (otherwise.) | 10 | 16 | 30 | 30 (>50 ch. [1]) 24 otherwise | 30 (>50 ch. [1]) 24 otherwise |
| Minimum Number of Hopping Channels | 50 (BW$^2$ < 250 kHz) 25 otherwise | - | - | - | - | 50 (BW$^2$ < 250 kHz) 35 otherwise | 50 (BW$^2$ < 250 kHz) 25 otherwise |
| Maximum Bandwidth of Hopping Channels (kHz) | 500 | - | - | - | - | 500 | 500 |
| Maximum Spurious Emission Threshold. (dBuV/m@3m) | 54 | 66 | 66 | 66 | 66 | 54 | 54 |
| **Parameters for Medium Access based on Duty Cycle** | | | | | | | |
| Band Duty Cycle (%) | - | 0.1 (863-868) 1 (865-868) 0.1 (868.7-869.2) 10 (869.4-869.6) 1 (870-875.6) | - | - | 1 | - | - |
| Band Duty Cycle Period (s) | - | 3600 | - | - | 3600 | - | - |
| Channel Duty Cycle (%) | 2 (BW2 < 250 kHz) 4 (250 kHz < BW$^2$ < 500 kHz) | - | - | - | - | 2 (BW$^2$ < 250 kHz) 4 (250 kHz < BW$^2$ < 500Hz) | 2 (BW$^2$ < 250 kHz) 4 (250 kHz < BW$^2$ < 500 kHz) |
| Channel Duty Cycle Period (s) | 20 (BW$^2$ < 250 kHz) 10 (250 Hz < BW$^2$ < 500 kHz) | - | - | - | - | 20 (BW$^2$ < 250 kHz) 10 (250 kHz < BW$^2$ < 500 kHz) | 20 (BW$^2$ < 250 kHz) 10 (250 Hz < BW$^2$ < 500 kHz) |
| **Parameters for Medium Access based on Polite Spectrum Access** | | | | | | | |
| Polite Spectrum Access Method | - | LBT[3], AFA[4] | - | - | - | - | - |
| Minimum Listening Time Window (μs) | - | 160 | - | 128 (SCS[5]) 5000 (LCS[6]) | - | - | - |
| Carrier Sense Level (dBm) | - | n.a.[7] | - | -80 | - | - | - |
| Minimum Toff (ms) | - | 100 | - | 2 (SCS5 if Tx-on > 6ms) 50 (LCS6) | - | - | - |
| Maximum Continuous Tx-On (s) | - | 1 (single[8]) 4 (dialoge[9]) | - | 0.4 (SCS5) 4 (LCS6) | 1 (single[8]) 4 (dialoge[9]) | - | - |
| Maximum Cummulative Tx-On | - | 100s/1h over 200 kHz of the spectrum | - | 360s/1h (SCS[5]) | 100s/1h over 200 kHz of the spectrum | - | - |

1- Number of hopping channels
2- Bandwidth of the hopping channels
3- Listen Before Talk, a medium access method where the transmitter avoids using the channel is it senses that someone is using the medium before the transmission
4- Adaptive Frequency Agility, a medium access method where the transmitter changes to another frequency channel if it detects that the current is being used. It can be used in conjunction with LBT.

5- Short Carrier Sense
6- Long Carrier Sense
7- Carrier Sense Level is not defined in Europe, some indications are given in ETSI TR 102 313 V1.1.1 (2004-07)
8- A single continuous transmission on a channel
9- A multiple transmissions as part of a bidirectional protocol

So regulator usually tries to reduce that probability by enforcing a **Medium Access Policy**. A possible policy is to enforce a **Band Tx Duty Cycle**. That is a percentage of time when the device can be actively emitting in the whole band. By doing so the regulator creates the opportunity that the channel is time multiplexed. If the endpoints would be perfectly coordinated, the number of potential transmitters would be inversely proportional to the duty cycle, but, in practice, there is no coordinator and collisions occur.

If no period is specified, there is a risk that transmitters take an arbitrary long cycle time as the denominator to compute the duty cycle. This would prevent others to use the channel for an undetermined period of time. To address this issue, the regulator can specify a **Band Tx Duty Cycle Period**, enforce a **Maximum Band Tx-ON Time** (the maximum time that a transmitter can be actively emitting continuously), or use both methods simultaneously.

Another possible policy is to enforce a **Polite Spectrum Access** mechanism such as Listen Before Talk (LBT) or Adaptive Frequency Agility (AFA). In case of LBT the regulator might specify a **listening time window**, and the minimum value of the signal strength above which is considered signal and not noise. This value is known as **Carrier Sense Level**. Polite policies can also enforce a maximum transmission time and a **Minimum Band Tx-OFF Time**, so that other transmitters have the chance to gain access to the medium.

In order to harmonize the use of the band, the regulator could enforce or restrict **modulation techniques** or the channelization of the band, i.e., **number of channels**, and **channel width**.

On channelized bands Frequency Hoping Spread Spectrum (FHSS) can be used. If the regulator allows this, it could specify different duty cycles for each of the sub-channels, while maintaining a global duty cycle for the band, or just removing the band restriction. This is usually done by specifying a maximum transmission time on the sub-channels, which is known as **Channel Dwell Time**. Optionally, it is also possible to specify a **Channel Duty Cycle**, and **Channel Duty Cycle Period**.

In such multichannel scenarios, the regulator could also decide to put a limit to the number of channels used, in other words, the **Total Used Bandwidth**.

Finally, there is a need to enforce transmitters to avoid spurious emissions significant to unintended frequencies out of the working frequency range, which could potentially affect transmitters on licensed bands. The regulator usually specifies a **Maximum spurious emission level** to prevent this from happening. Table 3 summarizes some of the most important values affecting the regulations for the higher frequencies of the unlicensed sub-GHz bands in main world markets. The first observation is that the allocated frequencies are different.

Other parameters also vary from country to country. In this context, and with the current globalization of the semiconductor industry, one can guess that this disparity of regulations does not benefit device manufacturers. We will later insist on this issue on Section X.

## VII. Maximum Node density for Spectrum use Worst Case Scenario

Because of the expected massive deployment of IoT technology, many studies analyze the potential maximum number of devices using a certain technology ([23][24]). However, those analyses are often not realistic because they underestimate the interference caused by other technologies working on the same unlicensed bands.

As we have previously mentioned sub GHZ ISM bands are extremely interesting for low power and long range networks since the required transmission power (as attenuation) has a quadratic relation with frequency (see Eq. 2). The low power scenario generally assumes that devices will have a good incentive to reduce the number of bytes transmitted to reduce energy consumption, since many will run on batteries. But this is not enforced. Nothing prevents devices connected to the mains power supply from using those ISM bands.

In this context, the Worst Case Scenario (WCS) analysis should ignore the minimum data transmission incentive and assume the maximum possible usage allowed by the regulator.

We would like to know the maximum number of devices transmitting on the unlicensed band in a certain area by assuming that they will try to work near the limits of regulation. Since receiving is not restricted by regulation, we will only consider transmission. The density of transmitters ($n_\rho$) will be defined by Eq. 3, where $n$ is the number of successful transmitters and $a$ is the area expressed in square kilometers. Thus, density of transmitters would be expressed in devices per square km ($dev/km^2$).

$$n_\rho = \frac{n}{a} \qquad (3)$$

If we are using a number of channels on the frequency band and a duty cycle, we can observe that the total number of devices in a certain area is given by Eq. 4, where $n_c$ is the simultaneous number of devices transmitting on the same channel, $r$ is the number of channels and $\propto$ is the duty cycle.

$$n = \frac{n_c r}{\propto} \qquad (4)$$

We can rewrite Eq. 3 as Eq. 5. and define $n_{c_\rho}$ as the density of nodes per channel.

$$n_\rho = \frac{n}{a} = \frac{n_c r}{a \propto} = n_{c_\rho} \frac{r}{\propto} \qquad (5)$$

Another important point in the IoT narrative is that it will produce a huge upstream traffic of real-time data coming from remote sensors to the Cloud. Downstream traffic is expected to be marginal. Collected data will be stored, mined, analyzed, and visualized using BigData and (lately) Deep-Learning algorithms. To understand how this goal can be achieved we propose to calculate the aggregated traffic density, i.e. aggregated network traffic per area, which would be expressed in bits per second per square kilometer ($bps/km^2$).

To compute the aggregated traffic density of the band, we should sum the network traffic of all the transmitters considering they are only transmitting during a duty cycle and divide them by the area, such as in Eq. 6, where $C_i$ is the capacity of the channel used by the transmitter $i$.

$$C_\rho = \frac{\sum_{i=1}^n \propto C_i}{a} = \frac{n \propto C_i}{a} = n_\rho \propto \quad C_i = \frac{n_c r\, C_i}{a} \quad (6)$$

It is interesting to realize that traffic density $(C_\rho)$ is independent of duty cycle.

As the maximum traffic on the channel should be the total band traffic capacity divided by the number of channels...

$$C = r\, C_i \quad (7)$$

...we can use (7) to rewrite (6) as (8)

$$C_\rho = \frac{n_c C}{a} \quad (8)$$

Again it is interesting to realize that the traffic density is also independent of the number of radio channels in which the band is split. So the main question again remains: what is the maximum number of simultaneous successful transmitters that can coexist in a certain area using the same channel $n_{c_\rho}$? But this question is ambiguous as we should define what a successful transmission is, and more important, where the receivers of the transmissions are located, as we known (from Eq. 2) that distance is a crucial factor for the receiving power.

*A. Scenario 1*

Imagine that all transmitters send to a single receiver and that they are located in a radius d from it, with enough radio coverage. Obviously, the number of simultaneous transmissions would be 1 ($n_c = 1$) and the area would be the coverage circle around the receiver. In this situation, the node density would be defined by Eq. 10.

$$n_\rho = \frac{n_c r}{a \propto} = n_{c_\rho} \frac{r}{\propto} = \frac{1}{\pi d^2} \frac{r}{\propto} \quad (10)$$

...and the traffic density by Eq. 11

$$C_\rho = \frac{\propto}{\pi d^2} \quad (11)$$

*B. Scenario 2*

Imagine an infinite number of transmitters randomly located in a square of $h \times h$ and that all receivers are placed in a coverage zone inside a circle of radius $d$ of its transmitter, such that $d \epsilon (0, h)$. In this case, we should use probability analysis to compute the maximum number of successful transmitters.

We denote $Tx_i$ as the event the transmitter $i$ being successful and $p_i$ its location. We consider that any transmitter closer to the distance d will interfere with our signal, making it to fail.

In this case, $P(Tx_0) = 1$ as the first transmitter, without the presence of anyother transmitter will always be able to transmit. A second transmitter will be able to transmit, only if it is located further from the first one by a threshold distance d. So $P(Tx_1) = P(|p_1 - p_0| > d)$. A third transmitter will be able to transmit, only if it is located further from the first and the second. So $P(Tx_2) = P(|p_2 - p_0| > d)\, P(|p_2 - p_1| > d)$.

Being $p_i$ a random variable, we can select another random variable $w$ which is the distance between two samples of p, and we can generalize the Eq. 12 for any transmitter.

$$P(Tx_i) = P(w > d)^i = (1 - P(w \le d))^i \quad (12)$$

Since $P(w \le d)$ is the cumulative distribution function of the random variable w, which can be rewritten as $CDF_w(d)$, we can count how many successful transmitters there are just by adding their probabilities of success, see Eq. 13.

$$n_c = \sum_{i=0}^\infty P(Tx_i) = \sum_{i=0}^\infty \left(1 - CDF_w(d)\right)^i$$

$$n_c = = \frac{1 - CDF_w(d)}{CDF_w(d)} \quad (13)$$

But we need to know the distribution function of the distance of two random points in space. Following an analysis similar to [25] and ignoring the corner cases we find the expression Eq. 14.

$$CDF_w(d) = \frac{d^4}{2h^2} - \frac{8}{3}\left(\frac{d^2}{h}\right)^{\frac{3}{2}} + \frac{\pi d^2}{h} \quad (14)$$

As the area of the square tends to infinite, the density is defined by Eq. 15.

$$\lim_{h \to \infty} n_{c_\rho} = \lim_{h \to \infty} \frac{n_c}{a} = \frac{1 - \left(\frac{d^4}{2h^2} - \frac{8}{3}\left(\frac{d^2}{h}\right)^{\frac{3}{2}} + \frac{\pi d^2}{h}\right)}{a\left(\frac{d^4}{2h^2} - \frac{8}{3}\left(\frac{d^2}{h}\right)^{\frac{3}{2}} + \frac{\pi d^2}{h}\right)} = \frac{1}{\pi d^2} \quad (15)$$

So, this results on exactly the same expressions for $n_{c_\rho}$ as in the first scenario.

In this analysis, we have used a value for the $d$ distance equal to the coverage radius of the transmitting and receiving endpoints. This value is often empirically found depending on the type of scenario (rural or urban), the frequency bands, and the modulation used.

From the regulation perspective, the regulator can try to control this radius by either specifying a maximum transmission power (since limiting the transmission power limits the range) or making listen before talk mandatory and specifying a carrier sense level.

## VIII. Theoretical Maximum Densities for LoRa and Sigfox

LoRa® [26] and Sigfox™ [27] are currently two popular IoT-LTN technologies. Both technologies are very different from each other and adapt to the regulatory landscape in different ways. Moreover, their proposers base their business models in a different part of the value chain.

LoRa is promoted by Semtech Corporation, who holds some patents parts on its physical channel (like [28]). It sells transceiver chips and IP to other semiconductor companies and integrators. The LoRa Alliance™ promotes the LoRaWAN™ networking protocol based on the LoRa physical layer. Users can deploy their own LoRa gateways and build their network, or possibly use existing infrastructure from other organizations. There are initiatives to create collaborative network infrastructure (such as The Things Network [29]), but there are also traditional many telecom operators providing the infrastructure.

On the other hand, Sigfox is promoted by the company with the same name. Sigfox also holds some patents on the physical layer, but their IP can be accessed freely by the members of the Sigfox consortium, so IP licensing is not the core of the business model. On the contrary, the company is focussed on deploying the network infrastructure at the global scale and offering it as a service.

From the technical point of view, the physical layers are very different. For this analysis we will only consider uplink channels and ignore downlink ones, since this is the factor that will limit the scalability of the system. LoRa uplink channels use a Chirp Spread Spectrum (CSS) modulation, or optionally Frequency Shift Keying (FSK). Obviously, this requires a significant bandwidth, so channels use either 125 kHz or 250 kHz in Europe and up to 500 kHz in the US (see [30]).

Sigfox uses a different approach based on Ultra Narrow Band (UNB) channels of 200 Hz with Binary Phase Shift Keying (BPSK).

Although the physical layers have a given traffic capacity, the LoRaWan Alliance and Sigfox Consortium limit themselves to a number of channels on certain frequencies to ensure interoperability.

The frequency plan of LoRa for different Regions is specified in [31] . LoRa specifies some standard channels for every region and allows the allocation of new channels dynamically based on applications. However, a gateway will usually have a limit on the number of channels that can be listening, so to have an estimation on the typical traffic capacity we might assume that only the standard channels are used.

Sigfox is using 360 channels, with a traffic capacity of 100 bps per channel in Europe and 600 bps in the US.

Table 4 and Table 5 show the calculated aggregated traffic capacity for different regions using LoRa and Sigfox technologies respectively .

Additionally, as we know from Eq. 2, the coverage radius depends on frequency and emitting power. Since regulation is different, and we have not found an empirical analysis of LoRa coverage in different countries, we obtain values for different countries applying the former equation and the maximum allowed transmission power starting with the assumption that a realistic coverage radius for a rural deployment in Europe is 10 km for LoRa and 20 km for Sigfox. Results are shown in Table 6.

TABLE 4
LoRa TYPICAL AGGREGATED CAPACITY

| Region | | Bandwidth (kHz) | Mod. | Num Channels | Max Channel Capacity (bps) | Total |
|---|---|---|---|---|---|---|
| Europe | | 125 | CSS | 7 | 5470 | 38290 |
| | | 250 | CSS | 1 | 11000 | 11000 |
| | | 125 | FSK | 1 | 50000 | 50000 |
| | Total | | | 9 | | 99290 |
| US / Canada | | 125 | CSS | 64 | 5470 | 350080 |
| | | 500 | CSS | 8 | 12500 | 100000 |
| | Total | | | 72 | | 450080 |
| China | | 125 | CSS | 6 | 5470 | 32820 |
| | Total | | | | | 32820 |
| India | | 125 | CSS | 3 | 5470 | 16410 |
| | Total | | | | | 16410 |

TABLE 5
SIGFOX TYPICAL AGGREGATED CAPACITY

| Region | Bandwidth (kHz) | Mod. | Num Channels | Max Channel Capacit (bps) | Total |
|---|---|---|---|---|---|
| Europe | 0.1 | D-BPSK | 360 | 100 | 36000 |
| US/Canada | 0.6 | D-BPSK | 360 | 600 | 60000 |

TABLE 6
ESTIMATED COVERAGE RADIUS FOR RURAL ENVIRONMENT ON DIFFERENT TECHNOLOGIES AND REGIONS

| Region | Tx Power (dBm) | Frequency (MHz) | Estimated LoRa radius (km) | Esimated Sigfox radius (km) |
|---|---|---|---|---|
| Europe | 16 | 868 | 10.0 | 20.0 |
| US/Canada | 30 | 915 | 47.5 | 95.0 |
| China | 12.5 | 780 | 7.4 | 100.4 |
| India | 30 | 866 | 50.2 | 14.8 |

Duty cycle limits must also be considered. For LoRa in US, Canada, it would be 100% as there are no duty cycle limits affecting the whole band. In Europe and India, the limit would be 1%; and 0.1% in China. For Sigfox, the duty cycle must be computed taking into account the daily limit of 140 messages of a maximum of 12 bytes payload per day, per device,

With all the collected information we can predict the densities for the different technologies in different regions (see Table 7).

TABLE 7
DENSITY ESTIMATIONS FOR RURAL DEPLOYMENTS ON DIFFERENT TECHNOLOGIES AND REGIONS

| Technology | $\alpha$ | R | d (km) | C (bps) | $n_\rho$ (dev/km2) | $C_\rho$ (bps/km2) |
|---|---|---|---|---|---|---|
| LoRa Europe | 1% | 9 | 10 | 99209 | 2.9 | 315 |
| LoRa US/Canada | 100% | 72 | 47.5 | 450080 | 0.01 | 63 |
| LoRa China | 0.1% | 6 | 7.4 | 32820 | 34.9 | 190 |
| LoRa India | 1% | 3 | 50.2 | 16410 | 0.04 | 2 |
| Sigfox Europe | 0.0004% | 360 | 20 | 36000 | 71619 | 28 |
| Sigfox US/Canada | 0.0003% | 360 | 95 | 60000 | 4232 | 2 |

Focusing on the network traffic capacity (see Figure 3), there is a big difference between different technologies and their performance on various regulation landscapes. Obviously, there is a clear inverse relation between distance and the number of bits per second that can be extracted from a certain area. The European version of LoRa is the technology that offers the higher traffic capacity, but it has the drawback of increasing the cost for gateway deployment. On the other extreme, the North American Sigfox offers the higher coverage but at the lowest bitrate density of less than 3 bps/km². In any case, the bitrate density is always below 1 kbps/km².

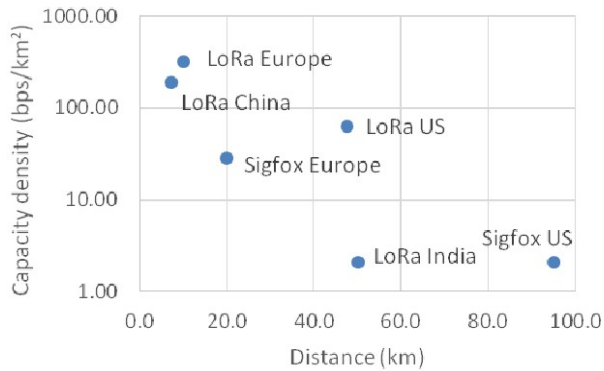

Figure 3 Traffic capacity density (in bits per second per square kilometer) of LoRa and Sigfox on the analyzed regulations. Traffic capacity density axis is plotted on a logarithmic scale.

The maximum device density (see Figure 4) is in the expected range for Sigfox but in a lower range for LoRa, especially in US and India, where the higher coverage radius, as a product of the higher allowed transmission power, goes against the device density. Sigfox high density is the result of their self-limitation on duty cycle, but it is important to understand that these numbers are ignoring the interference between technologies.

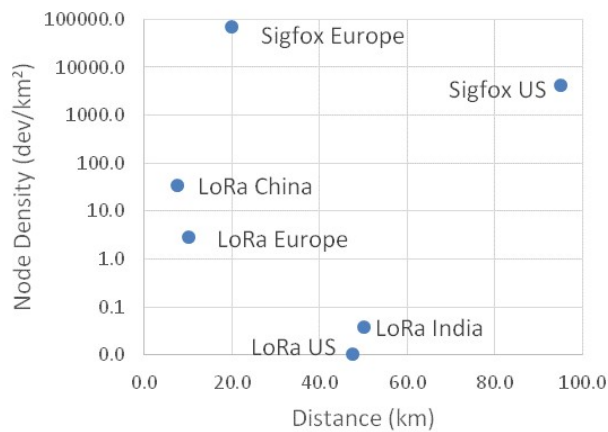

Figure 4 Maximum device densities of Lora and Sigfox networks assuming the maximum allowed transmission from end-devices and no interference from different technologies. Node density axis is plotted on a logarithmic scale.

## IX. STATE OF THE ART RESULTS ON DENSITIES

Previous theoretical analysis has a limited value. First, we might have been wrong in estimating the coverage radius, so we could have been wrong on the coverage capacity and, as a result, misestimated the node density for some scenarios. On the other hand, we have totally ignored interference, so, most probably, we have overestimated the capacity of the channel, an overestimated the node and capacity densities.

To shed some light on these issues we will have analyzed different density analysis in the literature. The different works are either based on analytical formulations, on simulation after the characterization of the fundamental communication properties, or on the analysis of real deployments.

### A. Analytical studies

In some analytical studies (like [24][32][33]) they analyze the deployment of an application with a certain traffic pattern. The difference with our analytical study is that they do not consider the worst case scenario imposed by regulation, but a more optimistic one. We analyze those works but only consider their reported successful transmissions. We try to harmonize the metrics so that we can compare all works.

Some works (like [24]) provide the message period ($T_{msg}$) between two consecutive packets from the same device, the message size in bytes ($S_{msg}$), and the number of successful transmitters ($n$), from which we can obtain an estimate of the total aggregate network traffic by Eq. 16.

$$C = \frac{n\,S_{msg}}{T_{msg}} \qquad (16)$$

Other works (such as [33]) provide the number of packets per hour per node ($f_{pph}$). By a simple conversion ($T_{msg} = 3600/f_{pph}$) we can find the message period and then apply Eq. (16) to get network traffic as Eq. (17).

$$C = \frac{n f_{pp}\,S_{msg}}{3600} \qquad (17)$$

Table 8 shows the calculated densities derived from the information of those works, which all use LoRa technology. Coverage radius for all works is below 10 km, which seems a little optimistic in the light of many previous coverage analyses. The resulting capacity density is generally below a few hundreds of bps and the device density is only above a few hundreds of devices when low activity is assumed.

TABLE 8
ANALYTICAL STUDIES

| | $T_{msg}$ (s) | $S_{msg}$ (B) | $n$ | d (km) | $C$ (bps) | $n_p$ (dev/km²) | $C_p$ (bps/km2) |
|---|---|---|---|---|---|---|---|
| [32](DR5, 3ch. scn.1) | 30 | 1 | 357 | 2.46 | 90 | 18.77 | 5 |
| [32](DR5, 3ch.scn.2) | 86400 | 8 | 842710 | 2.46 | 620 | 44325 | 32 |
| [32](DR1, 3ch. scn.3) | 600 | 20 | 335 | 7.32 | 80 | 1.99 | 0.5 |
| [24](Rd. Signs 6 ch.) | 30 | 1 | 8034 | 1.2 | 2140 | 1776 | 470 |
| [24](House apps. 6 ch.) | 86400 | 8 | 19444506 | 8.9 | 14400 | 78139 | 60 |
| [33](250 dev. 3 ch.) | 9.8 | 10 | 250 | 2.0 | 2040 | 19 | 160 |
| [33](5K dev. 3 ch.) | 200.0 | 10 | 5000 | 2.0 | 2000 | 397 | 160 |

### B. Simulation-based studies

Some analytical studies have the drawback of ignoring phenomena like bit error rate (BER) and interference. As shown in [34][35][37], the probability of packet loss increases as the number of nodes increases due to interference. The error probability has a direct relation with the time on air of the signal. It is known, in the case of LoRa, higher spreading factors increase time on air, causing more interference errors.

Simulation studies usually consider scenarios with a number of nodes ($n_{total}$) injecting an increasing number of packets to the network and reporting a probability (or rate) of packet transmission error ($P_{per}$) or probability of packet transmission success ($P_{psr} = 1 - P_{per}$).

For our analysis, we are going to use those probabilities (see Eq. 18) to obtain the effective number of successful transmitters.

$$n = n_{total}P_{psr} = n_{total}(1 - P_{per}) \qquad (18)$$

Again, different metrics are used to report the network traffic in the system, $f_{pph}$ in [34], $P_{per}$ in [35], and $P_{psr}$ in [36],[37],[38].

Most works use a quite realistic value for coverage radius of few km. In [35] they use a quite pessimistic value of 100m and [38] uses an optimistic 6 and 12 km scenario, while [37] uses the later. It is also strange how [36] does not locate the gateways on the center of the coverage areas.

In [38] there is no use of packet size, so we derive the capacity by using the specified duty cycle. In [37] neither packet size, nor duty cycle is specified.

TABLE 9
SIMULATION STUDIES

| | $T_{msg}$ (s) | $S_{msg}$ (B) | $n$ | d (km) | $C$ (bps) | $n_\rho$ (dev/km²) | $C_\rho$ (bps/km2) |
|---|---|---|---|---|---|---|---|
| [34] (1 channel) | 51.42 | 20 | 100 | 3.5 | 311 | 2.59 | 8 |
| [34] (3 channels) | 27.69 | 20 | 200 | 3.5 | 1156 | 5.19 | 30 |
| [35] | 1000 | 20 | 480 | 0.1 | 76 | 15278.87 | 2444 |
| [36] VSF Naville | 300 | 10 | 21 | 0.84 | 5 | 9.58 | 2 |
| [36]VSF Saragozza | 300 | 10 | 23 | 1.27 | 6 | 4.58 | 1 |
| [37] | | | 400 | 12 | | 0.88 | |
| [38] 6 km | | | 1100 | 6 | 14440 | 9.72 | 127 |
| [38] 12 km | | | 600 | 12 | 7876 | 1.32 | 17 |

The calculated densities for these works are shown in Table 9. The values for capacity density are consistently small, in a similar range than the previously found with analytical methods. An exception is the value from [35], which is the result of having a coverage radius of 100m. Such a small value seems unacceptable for the target applications of LPWAN.

Regarding the density of devices, most of the values are below 10 devices per km2, with the exception of the former case with an unacceptable coverage radius. Those values are significantly lower than the previously reported in analytical studies.

### C. Real deployments.

Real deployments are a better source of information, but due to the cost of deploying a large number of devices, some recent works limit themselves to a very small number of devices (like [34][35][39][40][41][42]) contributing few interesting information rather than realistic coverage measures in different scenarios.

On [43] they provide a slightly more realistic deployment on Congo, although with an extremely limited number of devices.

A more important deployment is described in [44]. They analyze data from the "The Things Network" over a period of 8 months, i.e. 21 Ms. They collected 17467312 packets with an average payload size of 18 bytes. This gives a total of 2.5 Gb traffic and an effective network capacity of 119 bps. The first thing to see here is that the network is heavily underused. In this case, the coverage area is not reported, but they report the number of gateways 691. By a conservative 1 km² coverage per gateway and taking into account the reported 1618 end devices we can find a realistic value for node and capacity densities.

The results from another deployment in Lyon is described in [45] containing 10 LoRa sensors and 4 Sigfox sensors. They report neither message size, nor message period. But, from the reported Daily Packet Loss statistics for LoRa sensors, we can obtain a message frequency of 50 packets per day per sensor and assume a message size of 8 bytes. From the same information, we can derive a $P_{psr}$ of 0.89.

In [46] the authors describe a deployment based on DQ-N, a technology based on LoRa transceivers. According to the paper a DQ-N gateway supports up to 5712 nodes generating an uplink traffic of 30 Bytes/hour with 36 Bytes packets. They do not report the exact coverage radius, but they suggest a 10 km typical coverage radius. In this case, no packet error probability is reported. Hence, results should be taken with a grain of salt.

TABLE 10
REAL DEPLOYMENTS

| | $T_{msg}$ (s) | $S_{msg}$ (B) | $n$ | d (km) | $C$ (bps) | $n_\rho$ (dev/km²) | $C_\rho$ (bps/km2) |
|---|---|---|---|---|---|---|---|
| [43](Congo Fridges) | 900 | 8? | 13.3 | 0.9 | 0.9 | 5.22 | 0.37 |
| [44] (TTN) | | | 1618 | 14.83 | 119 | 2.34 | 0.173 |
| [45] | 1728 | 8? | 10 | 1 | 0.37 | 2.86 | 0.117 |
| [46] | 4320 | 36 | 5712 | 10 | 380 | 18.18 | 1.21 |

Reported densities are very low, and lower than previous simulation results. Actually the effective duty cycle of the analyzed deployments is below 0.0015%, much below the regulation limits. With this parameters we can assume that not much interference is happening.

These values are another proof that there is a need for more research on real LTN network deployments and their scalability issues.

## X. THE ECONOMIC IMPACT

As detailed in [47], a typical IoT-LTN end device is an embedded device consisting of a processor connected to a radio transceiver, a number of sensors or actuators, a power supply, and the required volatile and non-volatile memory. Quite often, some of the parts can be included in the same chip. Low power embedded microcontrollers usually include the memory blocks, and there are radio transceiver SoCs integrating most of the components.

IoT-LTN gateways have a significantly higher cost due to their higher communication needs and computing power. They must receive and transmit data from several radio channels and connect with the Internet.

The cost of a deploying a $n$ number of end devices in a region of area $a$, is given by the Eq. 19. Where $X_{gw}$ is the cost of the gateway, $X_{dev}$ is the cost of an end device, $\propto$ is the duty cycle, $r$ the number of channels, and $d$ is the coverage radius of gateways. The equation is the result of assuming that the gateway density is $1/d$.

The $n$ devices system can only be deployed if the node density is lower than node density value defined by Eq. 10, which basically depends on the number of channels of the technology in use and the duty cycle. If the system cannot meet the required number of nodes, the cost is assumed to be infinite.

$$X = \begin{cases} a\frac{X_{gw}}{\pi d^2} + nX_{dev}, & \frac{n}{a} < n_\rho \\ \infty & , \text{ otherwise} \end{cases} \qquad (19)$$

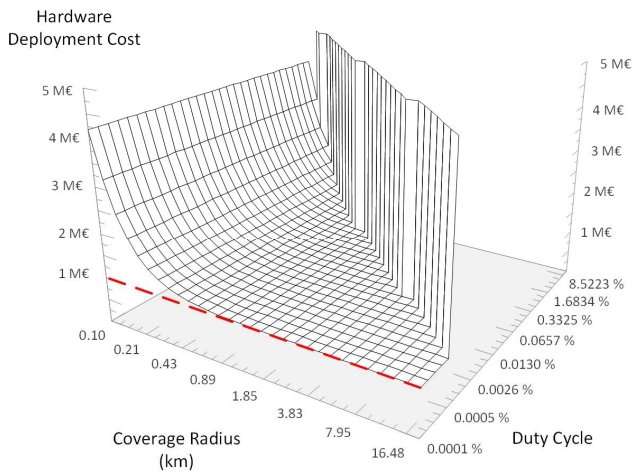

Figure 5 Hypothetical trade-off between duty-cycle and coverage radius for the cost of the hardware needed for the deployment of 100 k devices in a 100 km² area using LoRa with 8 channels. In this example we are assuming a cost of 10€ for the end device and 1000€ for the gateway device.

In this context, a system deployer should decide a system with a coverage radius that minimizes the cost, but ensuring that the device density is big enough to accommodate the expected number of devices. As detailed in [48], in the mid-term, the market expects a cost lower than 5$ for end-devices. Current prices are typically higher because of the reasons we will see in the following subsections. For the example illustrated in Figure 5 we have assumed an end device price of 10€. The example shows the total cost of the hardware for the deployment of 100 k devices on an area of 100 km², with different coverage radius and duty cycle. In this example the

factor between gateway and device costs is defined by $X_{gw} = 100X_{dev}$.

As the coverage radius is increased the number of gateways needed is reduced, the aggregated cost of the needed gateways is much lower than the aggregated cost of the required devices, and the total cost tends to the value of the devices. However, increasing the radius reduces the node density and requires the duty cycle to be reduced so that all devices can be accommodated.

## A. The risk for attacks

The drawbacks of using an unlicensed band is that you cannot prevent others from using the spectrum. Anyone could inject traffic to the air with malicious objectives. To the best of our knowledge, under current regulations, this would be totally legal.

The most basic attack could be the jamming of the radio channels. As seen in previous sections, coverage radius can be significant, and system developers must decrease the duty cycle to much lower levels than those allowed by regulations to build a successful implantation. A potential attacker could jam the radio channels with a small number of end devices working at the limits of the regulation. Taking into account the low cost of the devices, the risk seems high.

As detailed in [49][50], more elaborated attacks are possible, such as the replay of emitted packets.

## B. The Economics of Microelectronic Systems

As seen previously, the cost for an IoT system deployment should be dominated by the end device cost, so there is a big pressure for the device producers to decrease their manufacturing cost so that they can be massively produced and deployed.

The microelectronics industry is characterized by its economy of scale, so the more units of the same product you produce, the lower price you can achieve. On the other hand, different applications will need different hardware. There is a need of being able to reuse the same chips or modules for the broader possible scope.

The spectrum regulation current differences among countries make it harder for industry players to meet this goal. Radio transceivers manufacturers have adapted to the situation by covering a large spectrum and allowing configuration of many parameters of the radio link.

On the processor side, the software stack must be adapted to all the different behavior rules (like duty cycle requirements) that can be easier controlled by the higher levels of the communication stack.

A more difficult roadblock is the frequency of operation. The disparity on frequencies of country regulations makes the wavelength vary from 32 cm to 38 cm, or even 69 cm when we consider the 433 MHz band. This has an impact on the antenna selection (see [51]). In order to work in all scenarios, a high-bandwidth antenna should be used, but it would require a lot of space and its price is higher than many smaller alternatives.

On the other hand, lower cost antennas (like SMD ceramic, or PCB based antennas) work in a much smaller bandwidth, making it difficult to work on different regulations. A mismatch between the center frequencies in such antennas can produce a significant efficiency reduction. In [52] they empirically demonstrate how the coverage range of a LoRa system is reduced as much as 20% when using a PCB antenna designed for the 915 MHz band in the 868 MHz band.

Figure 6 illustrates different devices in the market at scale. It is obvious that the antenna is a limiting factor for the devices. A system with no antenna (a) can be as small as 12 × 13 mm. A ceramic antenna (b) is one of the smallest options followed by PCB based antenna (c). However, they have the drawback of the low bandwidth. On the other hand, external large antennas (d) increase the cost because of the more expensive antenna, the cost of the connectors, and the bigger mechanical requirements.

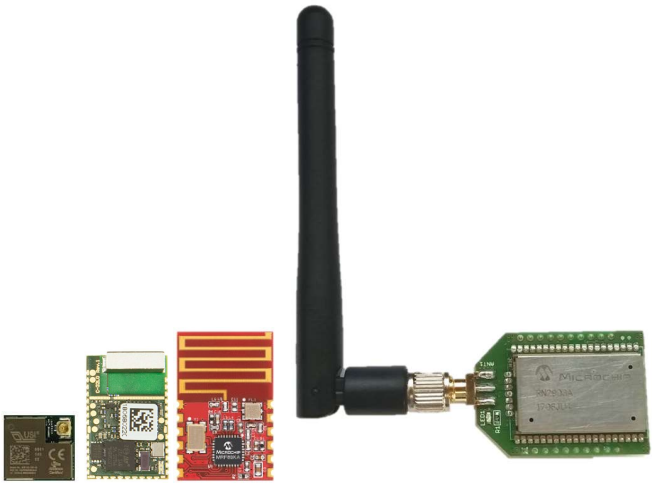

Figure 6 a) USI's LoRa transceiver USI WM-SG-SM-42 with a connector for an external antenna. b) Miromico FMLR-72-C-STL0 fully integrated LoRa sensor with a ceramic antenna. c) LoRa transceiver module based on the Microchip's MRF89XAM8A chip and a PCB antenna. d) LoRaBee Module RN2903 connected to an external antenna.

### C. The certification overhead

The current regulation situation requires a different certification for all countries. The management of the certification processes for global deployment is an overhead, but not much different than for other technologies working at the 2.4 GHz band.

However, the use of different frequencies can cause to create different devices for different regions. By having more than one single reference device, due to different hardware configurations, this may lead to multiple certification programs increasing the cost of testing and approval, impacting the final price of the device.

Trusting recognized testing laboratories and certification bodies is crucial for making sure that the device meets applicable regulations, confirming all related paperwork is up-to-date, and avoiding any potential market-surveillance issues due to non-compliances.

## XI. A REGULATION TO FOSTER IoT-LTN APPLICATIONS

It is clear that current regulations present a number of risks to the deployers of IoT systems using the unlicensed LTN bands. This is especially problematic for network operators providing connectivity based on them since the uncontrolled scenario makes it extremely hard to ensure any quality of service. It is safer to invest in technologies working in licensed bands like NB-IoT. Although the laws that limit the scalability are similar to the unlicensed case (see [53]), the market forces will put adequate incentives for the correct use of the spectrum while preventing the most basic security attacks. However, operators could have little incentives to deploy the infrastructure if very little revenue is expected, especially in rural areas. In addition, the shorter range of NB-IoT (as detailed in [54]) makes it less appropriate for the rural scenarios. Paradoxically, those scenarios are some where WSNs could benefit better from LTN connections.

Some technologies like Weightless-N, already anticipate the use of licensed bands, which seems a good strategy in terms of the chip provider, but it is useless if operators do not adopt it (see [55]).

Current LP-WAN technologies provide a reasonably good coverage but limit the number of potential devices when considering worst case scenarios. In the light of the findings of section IX, realistic device densities on current technologies are below 10 dev / km². The habitable land on the earth is approximately 130 M km². Filling the earth with IoT devices at such density factor, we would get 1.3 G devices and a total aggregated bandwidth of 130 Gbps. Even covering the whole world, this number of devices is far from the stated numbers on many optimistic forecasts.

Coverage, node density, and bitrate density can be scaled-up by using directional antennas in the gateways but this has a limit of just about one order of magnitude and increases the cost accordingly. Reducing the coverage radius also increases the cost of the infrastructure (as depicted in Figure 5) and, for short-range communications, there could be higher bandwidth alternatives competing with LPWAN.

Regarding network traffic capacity density, all realistic analyses give results below 1 kbps / km² of unreliable traffic. This automatically limits the kind of applications that can be based on such traffic.

Even with all the limitations, we think that unlicensed bands could still be a medium for massive IoT deployment if a coordinated action among regulators would be adopted to minimize the risks and foster its use.

We advocate changes in the following areas:

### A. A single worldwide frequency band

The frequency band disparity among different world markets is significant. As shown in section X, it has an impact on the microelectronics industry, and the certification process.

A harmonization of the band would be beneficial at the global scale. The global 2.4 GHz ISM band is a good example of the benefits of such a strategy.

A single band would improve the economies of scale of microelectronic chip manufacturers and allow the integration of smaller antennas. It would also eliminate the need to create product variations to serve different markets, thereby reducing manufacturing, testing, and certification costs. Thus, the dimensions of end nodes would become smaller, and their cost could be significantly reduced, inducing a faster adoption of the technology.

*B.    Much lower maximum duty cycles for uplink*

Even without considering real-world interferences, the maximum node density with the current regulations seems too low to justify the deployment of the infrastructure. As shown in Eq. 10, the node density is mainly determined by the duty cycle and the number of radio channels.

The features of Sigfox are especially adequate to allow a good node density, since the UNB modulation uses a high number of channels and they impose themselves a very low duty cycle of less than 0.0004%. This results on thousands of devices per square kilometer. Nevertheless, you cannot stop others to go to much higher duty cycles, degrading the whole network performance, and making those numbers difficult to achieve.

A solution would be that a low maximum duty cycle would be enforced by regulators so that high densities would be possible. The maximum duty cycle value should take into account the coverage radius (see Figure 5), which is mainly the result of the transmitting power limit. So, the regulator should propose a combination of maximum transmitting power and maximum duty cycle to address the minimum device density that policymakers would like to permit.

*C.    Spectrum Efficiency*

LoRa CSS and FSK channels have a spectrum efficiency of 0.04 bps/Hz and 0.4 bps/Hz while Sigfox D-BPSK has a spectrum efficiency of 1 bps/Hz. At first sight, UNB has a superior spectrum efficiency than CSS, but there is still not enough literature about the real response with interference to be conclusive about both technologies. Anyhow, the regulator could enforce, as it has done for other bands, modulations above a minimum spectrum efficiency threshold.

*D.    Attacker prevention*

The proposed tougher limits on duty cycle automatically would make it harder for attackers to "legally" jam the radio channels, as more devices are needed to do so. But security is a serious concern. Since the risks are more complex additional measures should be thought to improve the security of LTN networks.

*E.    Coordinated certification*

In order to help the industry to push the technological change and reduce the certification times and deployments, it is of relevant need the setup of Mutual Recognition Agreements between different National Authorities in order to avoid extra testing and certification costs.

## XII.    Conclusion

This paper has reviewed the state of the regulations and certification schemes affecting products that want to gain access to the sub-GHz unlicensed band on the main world markets.    After an analysis of the implications of those regulations for the IoT systems, we have identified the main problem being the low density of end devices derived from the maximum allowed regulations parameters, such as the duty cycle. We have identified additional problems like the risk of security attacks that can hinder the business models of operators and system deployers.

We insist on the need to adapt and harmonize global regulations to boost the deployment of IoT so that its expected disruptive widespread becomes an industrial reality. Very recently, with Commission Implementing Decision (EU) 2018/1538 of 11 October 2018, the European Commision has adopted a decision to allocate a new unlicensed frequency band in 915-921 MHz. This is a good step towards the global harmonization of unlicensed bands that could foster LPWAN-based IoT,

In addition, the technical limitations set by regulators, and the differences between certification schemes have also been covered.    The present paper also gives manufacturers, importers and general players of IoT-LTN products an overview of the current requirements for accessing some of the most relevant markets worldwide.

## Acknowledgment

This work has been partly funded by the Serene-IoT project (Penta 16004), and had inputs from 3DSafeguard (ITEA 14034), and PEM-GASOL projects.

We would like to thank Dr. Josep Parrón Granados for his valuable comments on antenna design, Dr. Laura Prat Baiget for her help on probability functions, and Borja Herranz for his comments on commercial IoT modules.

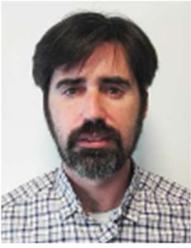

**David Castells-Rufas** was born in Manresa, Spain, in 1971. He received the B.S., M.S., and Ph.D. degrees in Computer Science from Universitat Autonoma de Barcelona (UAB), Spain in 1994, 2009, and 2016 respectively.

From 2003 he is working as researcher in the CEPHIS research group from the Microelectronics and Electronic Systems Department of the UAB, where he also teaches as a lecturer. He founded the technology based companies Histeresys and Creanium in 1998 and 2001 respectively.

His research interests include reconfigurable systems, high performance embedded systems, and computing architectures.

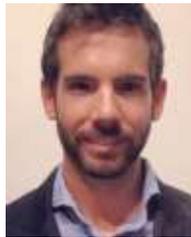

**Adria Galin** was born in Barcelona, Spain, in 1993. He received the B.S. degree and M.S degree in Telecommunication Systems Engineering in 2015 and 2017 respectively from Universitat Autonoma de Barcelona (UAB). Currently he is working as a Radio-Frequency engineer in the R&D department of Applus+ Laboratories. He is also member of REDCA, TCBC, ETSI and 3GPP with active participations and publications in several Working Groups. He also teaches as lecturer at the Universitat Politecnica de Catalunya (UPC). Main interests include real-time instantaneous frequency estimators, testing challenges of C-V2X and 5G, regulatory compliance and interference suppression receiver techniques

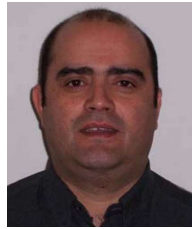

**Jordi Carrabina** was born in Manresa, Catalonia in 1963. He graduated in physics from the University Autonoma of Barcelona (UAB), Catalonia, Spain, in 1986, and received the M.S. and Ph.D. degrees in Microelectronics from the Computer Science Program, UAB, in 1988 and 1991, respectively. In 1986, he joined the National Center for Microelectronics (CNM-CSIC), where he was collaborating until 1996. Since 1990, he has been an Associate Professor with the Department of Computer Science, UAB. In 2005, he joined the new Microelectronics and Electronic Systems Department, heading the CEPHIS research group. Since 2004, CEPHIS has been recognized as TECNIO Innovation Technology Center from the Catalan Government Agency ACCIO. He is currently teaching in B.S. and M.Sc. Degree of Telecommunications Engineering and Computer Engineering at UAB, and the Masters of Embedded Systems at UPV-EHU and coordinating the New MsC Degree on IoT for eHealth. During last five years, he has coauthored more than 30 papers in journals and conferences. He has been a consultant for different international small and medium enterprises (SMEs) companies. His main interests are microelectronic systems oriented to embedded platforms, SoC/NoC architectures and printed microelectronics